\documentstyle[11pt,newpasp,twoside,psfig]{article}
\begin{document}
\title{Non radial motions and the large scale structure of the Universe}
\author{A. Del Popolo}
\affil{Dipartimento di Matematica, Universit\`{a} Statale di Bergamo,
Piazza Rosate, 2 - I 24129 Bergamo, Italy}
\author{ M. Gambera}
\affil{Istituto di Astronomia dell'Universit\`a di Catania, 
Viale A.Doria, 6 - I 95125 Catania, Italy}


\begin{abstract}
We study the effect of non-radial motions on the mass function, the VDF
and on the shape of clusters of galaxies using
the model introduced in Del Popolo \& Gambera (1998a,b; 1999). The mass
function of clusters, obtained using the quoted model,
is compared with the statistical
data by Bahcall \& Cen (1992) and Girardi et al. (1998),
while the VDF is compared with the CfA data by 
Zabludoff et al. (1993) for local clusters. In both cases the model
predictions are in good agreement with the observational data showing
once more how non-radial motions can reduce many of the discrepancies
between CDM model predictions and observational data. Besides we study
the effect of non-radial motions on the intrinsic shape of clusters of
galaxies showing that non-radial motions produce clusters less elongated
with respect to CDM model.
\end{abstract}


\section{Introduction}

At its appearance the CDM model contributed to obtain a better
understanding of the origin and evolution of the large scale structure
in the Universe (White et al. 1987).
The principal assumptions of the {\it standard} CDM (SCDM) model
(Liddle \& Lyth 1993) are:
\begin{itemize}
\item a flat Universe dominated by weakly
interacting elementary particles having low velocity dispersion at early
times. The barionic content is determined by the standard big bang
nucleosynthesis model; 
\item critical matter density; 
\item expansion rate given by $ h = 0.5$; 
\item a scale invariant and adiabatic spectrum with a spectral index, 
$ n \equiv 1$; 
\item the condition required by observations, that the fluctuations
in galaxy distribution, $(\delta \rho / \rho)_{\rm g}$ , are larger than the
fluctuations in the mass distribution, $(\delta \rho / \rho)_{\rho} $ by 
a factor $ b > 1$. 
\end{itemize}
If this last assumption is not introduced, the pairwise 
velocity dispersion is larger then that deduced
from observations and the galaxy correlation function is steeper than that
observed (Davis et al. 1985). After the great success of the model in the 80's, a closer inspection of the model has shown a series of deficiencies, namely:
\begin{itemize}
\item the strong clustering of rich clusters of galaxies, 
$ \xi_{\rm cc}(r) \simeq (r / 25h^{-1}{\rm Mpc})^{-2}$, 
far in excess of CDM predictions (Bahcall \& Soneira 1983);
\item the overproduction of clusters abundance. 
Clusters abundance is a useful test for models of galaxy formation. This
is connected to three relevant parameters: the mass function,
the VDF and the temperature function.
Using N-body simulations, Jing et al. (1994) studied the mass function
of rich clusters at $z=0$ for the CDM model. They 
found that the CDM model with the COBE normalization
produces a temperature function of clusters higher than that given by
the observations by Edge et al. (1990) and by Henry \& Arnaud (1991);
\item the
conflict between the normalization of the spectrum of the perturbation which
is required by different types of observations; 
\item 
the incorrect scale dependence of the galaxy correlation
function, $\xi (r)$, on scales
$10 \div 100$ $h^{-1} {\rm Mpc}$, having $\xi (r)$ too
little power on the large scales compared to the power on smaller scales
(Maddox et al. 1990; Saunders et al. 1991; Lahav et al. 1989;
Peacock 1991; Peacock \& Nicholson 1991).
\end{itemize}
Several alternative models have been proposed in order to solve
the quoted problems (Peebles 1984;  
Valdarnini \& Bonometto 1985; Bond et al. 1988;  
Holtzman 1989; Efstathiou et al. 1990; Turner 1991; 
Schaefer 1991; Schaefer \& Shafi 1993; Holtzman \& Primack 1993; Bower et al. 1993). 
Most of them propose 
in some way a modification of the primeval spectrum of perturbations.  
In two previous papers (Del Popolo \& Gambera 1998a; 1999) we showed
how, starting from a CDM spectrum and taking into account non-radial
motions, at least the problem of the clustering of clusters of galaxies
and the problem of the X-ray temperature can be considerably reduced. \\
Here, we study the effect of non-radial motions on the mass function, the
VDF and on the shapes of galaxy clusters.\\
In Sect. ~2 we shall use the same model introduced by
Del Popolo \& Gambera 1998a,b,1999) to compare the mass function
calculated using the CDM model, taking into account non-radial motions,
with the observed mass function obtained by Bahcall \& Cen (1992).
Then, we repeat the calculation for the VDF and compare the theoretical VDF
with the CfA data by Zabludoff et al. (1993). 
In Sect. ~4 we study the effect of non-radial motions
on the ellipticity of clusters and finally in Sect. ~5 we give
our conclusions.

\section{The mass function and the velocity dispersion function}

One of the most important constraints that a model for large-scale
structure must overcome is that of predicting the correct number density of
clusters of galaxies. This constraint is crucial for several reasons
(Peebles 1993). The abundance of clusters of galaxies, 
together with the mass distributions in galaxy halos and in rich clusters
of galaxies, the peculiar motions of galaxies, the spatial
structure of the microwave background radiation is one of the most
readily accessible observables which probes the mass distribution
directly.
The most accurate way of assessing the cluster abundance is via numerical
simulations. However, there is an excellent analytic alternative, Press \&
Schechter's theory (Press \& Schechter 1974; Bond et al. 1991).
Press-Schechter's theory states that the
fraction of mass in gravitationally bound systems larger than a mass, $M$,
is given by the fraction of space in which the linearly evolved
density field, smoothed on the mass scale $M$, exceeds a threshold
$\delta_{\rm c}$:
\begin{equation}
F(>M)=\frac{1}{2} erfc \left(\frac{\delta_{\rm c}}{\sqrt{2}
\sigma(R_f,z)}   \right)
\label{eq:press}
\end{equation}
where $R_{\rm f}$ is the comoving linear scale associated with $M$.
Press-Schechter's result predicts that only half of the mass of the Universe
ends up in virialized objects but in particular cases this problem can
be solved (Peacock \& Heavens 1990; Cole 1991; Blanchard et al. 1992). \\
The mass variance present in Eq.~(\ref{eq:press})
can be obtained once a spectrum, $P(k)$, is fixed:
\begin{equation}
\sigma ^2(M)=\frac 1{2\pi ^2}\int_0^\infty {\rm d}kk^2P(k)W^2(kR) 
\label{eq:ma3}
\end{equation}
where $W(kR)$ is a top-hat smoothing function:
\begin{equation}
W(kR)=\frac 3{\left( kR\right) ^3}\left( \sin kR-kR\cos kR\right) 
\label{eq:ma4}
\end{equation}
and the power spectrum $P(k) \propto k^n T^2(k)$ is fixed giving the
transfer function $T(k)$(here, we adopt that by Klypin et al. 1993),
where $k$ is the wave-number measured in units of ${\rm Mpc}^{-1}$.
This spectrum is valid for $k< 30 {\rm Mpc}^{-1}$ and $z<25$. The accuracy
of the spectrum is $5\%$. It is more accurate than Holtzman's (1989) spectrum,
used by Jing et al. (1994) and Bartlett \& Silk (1993) to calculate
the mass function and the X-ray temperature function of clusters, respectively.
The spectrum is lower by $20\%$ on cluster mass scales than 
Holtzman's (1989). The spectrum was normalized to the COBE quadrupole
$Q_2=17 \mu$ K, corresponding to
$\sigma_8=0.66$.
As shown by Bartlett \& Silk (1993) the X-ray distribution
function, obtained using a standard CDM spectrum, over-produces the clusters
abundances data obtained  by Henry \& Arnaud (1991) and by Edge et al.
(1990).
This has lead some authors (White et al. 1993)
to cite the cluster abundance
as one of the strongest pieces of evidence against the standard
CDM model when the model is normalized so as to reproduce the
microwave background anisotropies as seen by the COBE satellite
(Bennett et al. 1996). \\
The discrepancy can be reduced taking into account the non-radial
motions that originate when a cluster reaches the non-linear regime as
follows. A fundamental role in Press-Schechter's theory 
is played by the value of $\delta_{\rm c}$. Using a top-hat window function
$\delta_{\rm c}=1.7 \pm 0.1$ while for a Gaussian window the threshold is
significantly lower. In a non-spherical context the situation is more
complicated. Considering the collapse along all the three axes the threshold
is higher, whereas the collapse along the first axis (pancake
formation) or the first two axes (filament formation) corresponds
to a lower threshold (Monaco 1995). The threshold, $\delta_{\rm c}$,
does not depend on the background cosmology. \\
As shown by Del Popolo \& Gambera (1998a; 1999), if non-radial
motions are taken into account, the threshold $\delta_{\rm c}$
is not constant but is function of mass, $M$ (Del Popolo 
\& Gambera 1998a; 1999):
\begin{equation}
\delta _{\rm c}(\nu )=\delta _{\rm co}\left[ 1+
\int_{r_{\rm i}}^{r_{\rm ta}}  \frac{r_{\rm ta} L^2 \cdot {\rm d}r}{G M^3 r^3}%
\right]
\label{eq:ma7} 
\end{equation}
where $\delta _{\rm co}=1.68$ is the critical threshold for a spherical model,
$r_{\rm i}$ is the initial radius, $r_{\rm ta}$ is the turn-around radius and
$L$ the angular momentum.
In terms of the Hubble constant, $H_0$, the density parameter at current
epoch, $\Omega_0$, the expansion parameter $a$ and
the mean fractional density excess inside a shell of a given radius,
$\overline{\delta}$ Eq. (\ref{eq:ma7}) can be written as (Del Popolo 
\& Gambera 1998a; 1999):
\begin{equation}
\delta _{\rm c}(\nu )=\delta _{\rm co}\left[ 1+\frac{8G^2}{\Omega
_o^3H_0^6r_{\rm i}^{10}\overline{\delta} (1+\overline{\delta} )^3}
\int_{a_{\rm min}}^{a_{\rm max }}\frac{L^2 \cdot {\rm d}a}{a^3}%
\right]
\label{eq:ma8} 
\end{equation}
where $a$ is the expansion parameter, and $a_{\rm min}$ its value
corresponding to $r_{\rm i}$. The mass dependence of the threshold parameter,
$ \delta_{\rm c}(\nu)$, was obtained in the same way sketched in
Del Popolo \& Gambera (1999).
\begin{figure}[ht]
\psfig{file=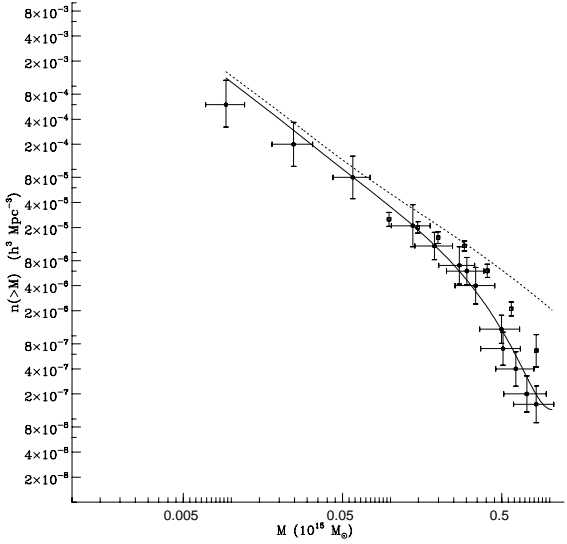,width=8.cm}
\caption{Cumulative mass function calculated using a CDM model without
taking into account non-radial motions (dashed line) and taking account
non-radial motions (solid line) compared with Bahcall \& Cen (1992) data (full dots)
and with that of Girardi et al. (1998) (open squares).}
\end{figure}
The result of the calculation is shown in Fig. 1.
Here, the mass function of clusters, derived using
a CDM model with $\Omega_0=1$, $h=1/2$ normalized to
$Q_{\rm COBE} = 17 \mu$ K and taking into account
non-radial motions (solid line), is compared with 
the statistical data by Bahcall \& Cen (1992) (full dots)
and with that of Girardi et al. (1998) (open squares)
and with a pure CDM model with $\Omega_0=1$, $h=1/2$ (dashed line).
Bahcall \& Cen (1992) estimated the cluster mass function using optical
data (richness, velocities, luminosity function of galaxies in clusters) 
as well as X-ray data (temperature distribution function of clusters).
Groups poorer than Abell clusters have also been included thus
extending the mass function to lower masses than the richer Abell clusters. 
Girardi et al. (1998) data are obtained from a sample of 152 nearby ($z \le 0.15$)
Abell-ACO clusters. As shown, the CDM model that does not take account of
the non-radial motions over-produces the clusters abundance.
The introduction of non-radial motions (solid line) reduces remarkably the
abundance of clusters with the result that the model
predictions are in good agreement with the observational data.
This result confirms what found in Del Popolo \& Gambera (1999) showing
how a mass dependent threshold, $\delta_{\rm c}(M)$,
(dependence caused by the developing of non-radial motions) can solve several of
SCDM discrepancies with observations.\\

The VDF is defined in a similar way
to the mass function, namely it is the number of objects per unit volume 
with velocity dispersion larger than $\sigma_{\rm v}$. Since the velocity 
dispersion $\sigma_{\rm v}$ can be observed directly, VDF provides a good 
test of theoretical models. Observed $\sigma_{\rm v}$ comes from the measurement of
galaxy redshift. The VDF can be calculated starting from the mass function:
\begin{equation}
n(\sigma_{\rm v}) =n(M) \frac{{\rm d} M}{{\rm d} \sigma_{\rm v}}
\label{eq:der}
\end{equation} 
The cumulative VDF can be obtained integrating Eq. (\ref{eq:der}):
\begin{equation}
n(>\sigma_{\rm v}) = \int_{\sigma_{\rm v}}^\infty n(\sigma_{\rm v'})
{\rm d} \sigma_{\rm v'}
\label{eq:derr}
\end{equation} 
In order to use Eq. (\ref{eq:der}) to calculate the VDF we need a 
relation between the velocity dispersion, $\sigma_{\rm v}$, and mass, $M$.
To determine the relation between $\sigma_{\rm v}$ and $M$ we use both
the relation for the typical virial temperature and the result by
Thomas \& Couchman (1992) and that by Evrard (1989, 1990, 1997)
found in N-body simulations. 
We find that the necessary relation between $\sigma_v$ and $M$ is given by:
\begin{equation}
\sigma_{\rm v}=824 {\rm km/s} \left(\frac{h M}{10^{15} M_{\odot}}\right)^{1/3}
\label{eq:ev}
\end{equation}
(Evrard 1989; Lilje 1990). 
\begin{figure}[ht]
\psfig{file=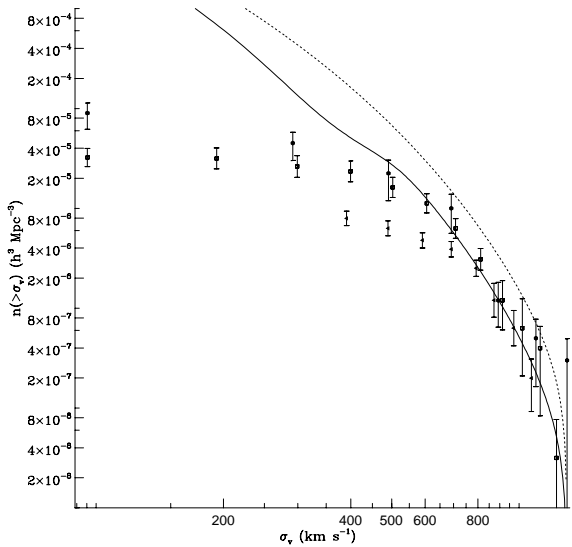,width=8.cm}
\caption{Cumulative VDF calculated using a CDM model without
taking into account non-radial motions (dashed line) and taking into account
non-radial motions (solid line) compared with Zabludoff et al. (1993) data (full dots)
and
with those by Mazure et al.(1996) (full triangles) for
$R \ge 1$ clusters and Fadda et al. (1996)
(open squares) for $R \ge -1$ clusters.
The theoretical curves are obtained using a $\sigma_{\rm v}$-$M$ relation with
zero scatter.}
\end{figure}
In Fig. 2 we compare the VDF obtained from a CDM model taking
account of non-radial motions (solid line) with the CfA 
data by Zabludoff et al. (1993) (full dots)
based on their survey of $R \geq 1$
Abell clusters within $z \leq 0.05$ and with the data by
Mazure et al. (1996) (full triangles) and Fadda et al. (1996) (open squares).
Mazure et al. (1996) data are obtained from a volume-limited
sample of 128 $R_{\rm ACO} \ge 1$ clusters
while that of Fadda et al. (1996)
are obtained from a sample of 172 nearby galaxy
clusters ($z \le 0.15$). 
We also plot the
VDF obtained from a CDM model without non-radial motions (dashed line). 
The SCDM model predicts more clusters than the CfA observation 
except at $\sigma_{\rm v} \simeq 1100  {\rm km/s}$. As reported by 
Jing \& Fang (1994) the SCDM model can be rejected at a very high 
confidence level ($>6 \sigma$). 
The reduction of
the normalization reduces the formation of clusters, thus resolving 
the problem of too many clusters, but leads to a deficit at 
$\sigma_v \simeq 1100  {\rm km/s}$.
When non-radial motions are taken into account (solid line) we obtain
a good agreement between theoretical predictions and observations. 
Both CDM and CDM with non-radial motions
predict more clusters of low velocity dispersion 
($\sigma_v \leq 300  {\rm km/s}$) than the observation. This discrepancy 
is not significant because the data at $\sigma_{\rm v}\leq 300 {\rm km/s}$ 
could be seriuosly underestimated (Zabludoff et al. 1993; Fadda et al. 1996;  
Mazure et al. 1996).
 
\section{Non-radial motions and the shape of clusters}

Most clusters, like elliptical galaxies, are not spherical and their
shape is not due to rotation (Rood et al. 1972; Dressler 1981).
The perturbations that gave rise to the formation of
clusters of galaxies are alike to have been initially aspherical 
(Barrow \& Silk 1981; Bardeen et al. 1986) and
asphericities are then amplified during gravitational collapse
(Icke 1973; Barrow \& Silk 1981). The elongations are
probably due to a velocity anisotropy of the galaxies
(Aarseth \& Binney 1978) and according to Binney \& Silk (1979) and to 
Salvador-Sol\'e \& Solanes (1993) the elongation of clusters originates
in the tidal distortion by neighboring protoclusters. In particular
Salvador-Sol\'e \& Solanes (1993) found that the main distorsion on
a cluster is produced by the nearest neighboring cluster having more than 45
galaxies and the same model can explain the alignement between neighboring
clusters (Oort 1983; Plionis 1993) and that between clusters and
their first ranked galaxy (Rhee \& Katgert 1987; van Kampen \& Rhee 1990;
West 1994).\\ 
The observational information on the
distribution of clusters shapes is sometimes conflicting.
Rhee et al. (1989) found that most clusters are
nearly spherical with ellipticities distribution having a peak
at $\epsilon \simeq 0.15$ while Plionis et al. (1991) found that
clusters are more elongated with the peak of the ellipicities
distribution at $\epsilon \simeq 0.5$.\\
To study the effect of non-radial motions on the shape of clusters
we shall use a model introduced by Binney \& Silk (1979).
In that paper they showed that tidal interactions between protoclusters and the
neighbouring protostructures should yield prolate shapes (before virialization)
with an
axial ratio of protostructures of $\simeq 0.5 $, the typical value found in
clusters. After virialization the pre-existing elongation is damped
and the axial ratio leads to values of about
$0.7 \div 0.8$, that are higher with respect to observations. As observed by
Salvador-Sol\'e \& Solanes (1993) this last discrepancy  can be
removed taking into account that tidal interaction keeps going on after
virialization and that on average the damping of elongations
due to violent relaxation is eliminated by its growth after
virialization. Then this growth restores a value of $\epsilon$
near the one that clusters had before virialization.\\
\begin{figure}
\psfig{file=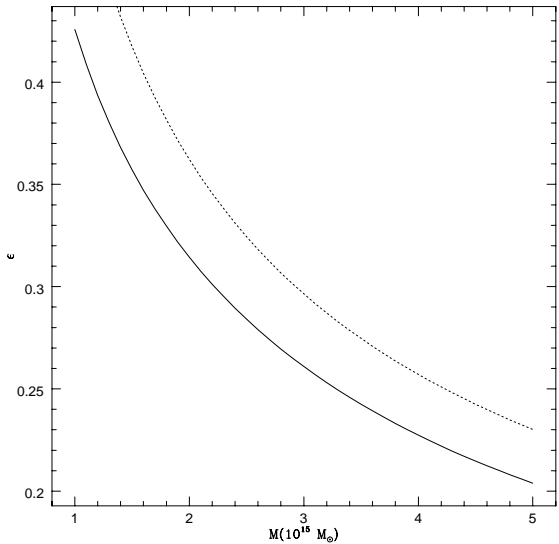,width=8.0cm}
\caption{Ellipticity, $\epsilon$, of clusters versus mass, $M$. The dashed
and solid lines
represent $\epsilon$ for a CDM without and with non-radial motions,
respectively.}
\end{figure}
According to the quoted Binney \& Silk (1979) model, an initially
spherical protostructure (e.g. a protocluster)
of mass $M$ having at distance $r(t)$ from its
centre a series of similar protostructure of mass $M'$ 
shall be distorted and the final ellipticity can be calculated
using the following equation:
\begin{equation}
\epsilon \simeq \frac{3}{2} \langle \mu^2 \rangle^{1/2} G(2,x_1)
\end{equation}
where $G(2,x_1)$ is defined in the quoted paper (see Eq. 13c) and
$\langle \mu^2 \rangle^{1/2}$ for a CDM without non-radial motions
is given by (Binney \& Silk 1979) while for a CDM with non-radial motions
we use the following relation:
\begin{equation}
\langle \mu^2 \rangle^{1/2} = \frac{1}{9 \rho_{\rm b}^{1/2}}
\left[\int \frac{N(M',r) {\rm d}M'}{M+M'}\right]^{1/2}
\end{equation}
In Fig. 3 we show as $\epsilon$
declines with mass in agreement with the observations. Besides,
we show as in a CDM model that takes
into account non-radial motions (solid line) $\epsilon$ is smaller than
in the simple CDM (dashed line). This show as
non-radial motions reduce the elongation of clusters. For a cluster of
$10^{15} M_{\odot}$ we get a value of $\epsilon \simeq 0.5$, if non-radial motions
are excluded, while $\epsilon \simeq 0.43$ when non-radial motions are taken
into account. Increasing the mass, as expected, clusters tend to become
more and more spherical. 

\section{Conclusions}

Here, we have studied how non-radial motions change the mass function,
the VDF and the shape of clusters of galaxies, using the model introduced
by Del Popolo \& Gambera (1998a; 1999).
We compared the theoretical mass function
obtained from the CDM model taking into account non-radial motions
with the experimental data by Bahcall \& Cen (1992)
and Girardi et al. (1998). The VDF was compared with the CfA data by 
Zabloudoff et al. (1993), by Mazure et al. (1996) and by 
Fadda et al. (1996). 
Taking account of
non-radial motions we obtained a notheworthy reduction of
the discrepancies between the CDM predicted 
mass function, the VDF and the observations. Non-radial motions are also
able to change the shape of clusters of galaxies reducing their
elongations with respect to the prediction of CDM.

\acknowledgements
We are grateful to Proff. G. Moncada, E. Recami and E. Spedicato 
for stimulating discussions while this work was in process.


\begin{references}
\reference Aarseth S.J., Binney J., 1978, MNRAS 185, 227
\reference Bahcall N.A., Cen Y., 1992, ApJ 398, L81
\reference Bardeen J.M., Bond J.R., Kaiser N., Szalay A.S., 1986, ApJ 304, 15 
\reference Barrow, J.D., Silk, J., 1981, ApJ 250, 432
\reference Bennet C.L., Banday A., Gorski K.M., et al., 1996, ApJ 464, L1
\reference Binney J., Silk J., 1979, MNRAS 188, 273
\reference Blanchard A., Valls-Gabaud D., Mamon G., 1992, A\&A 264, 365
\reference Bond J.R., Szalay A.S., Silk J., 1988, ApJ 324, 627
\reference Bond J.R., Cole S., Efstathiou G., Kaiser N., 1991, ApJ 379, 440
\reference Bower R.G., Coles P., Frenk C.S., White S.D.M., 1993, ApJ 405, 403
\reference Cole S., 1991, ApJ 367, 45
\reference Davis M., Efstathiou G., Frenk C.S., White S.D.M., 1985, ApJ 292, 371
\reference Del Popolo A., Gambera M., 1998a, A\&A 337, 96 
\reference Del Popolo A., Gambera M., 1998b, Proceedings of the VIII Conference on Theoretical Physics: General Relativity and Gravitation eds Vulcanov D.N. - Bistritza -  June 15-18, 1998 - Rumania
\reference Del Popolo A., Gambera M., 1999, A\&A 344, 17 (see also SISSA preprint astro-ph/9806044)
\reference Dressler A., 1978, ApJ 243, 26
\reference Edge A.C., Stewart G.C., Fabian A.C., Arnaud K.A., 1990, MNRAS 245, 559
\reference Efstathiou G., Sutherland W.J., Maddox S.J., 1990, Nat., 348, 705
\reference Evrard A.E., 1989, ApJ 341, L71
\reference Evrard A.E., 1990, ApJ 363, 349
\reference Evrard A.E., 1997, Sissa Preprint astro-ph/9701148
\reference Fadda D., Girardi M., Giuricin G., Mardirossian F., Mezzetti M., 1996, preprint SISSA astro-ph/9606098
\reference Girardi M., Borgani S., Giuricin G., Mardirossian F., Mezzetti M., 1998, preprint SISSA astro-ph/9804188
\reference Henry J.P., Arnaud K.A., 1991, ApJ 372,410
\reference Holtzman J., 1989, ApJS, 71, 1
\reference Holtzman J., Primack J., 1993, Phys. Rev. D43, 3155
\reference Icke V., 1973, A\&A 27, 1
\reference Jing Y.P., Fang L.Z., 1994, Phys. Rev. Lett. 73, 1882
\reference Jing Y.P., Mo H.J., Boerner G., Fang L.Z., 1994, A\&A 284, 703
\reference Lahav O., Edge A., Fabian A.C., Putney A., 1989, MNRAS 238, 881
\reference Liddle A.R., Lyth D.H., 1993, Phys. Rev., 231, n 1, 2
\reference Lilje P.B., 1990, ApJ 351, 1
\reference Maddox S.J., Efstathiou G., Sutherland W.J., Loveday J., 1990, MNRAS 242, 43p
\reference Mazure A., Katgert P., den Hartog R., et al. 1996, A\&A 311, 95 
\reference Monaco P., 1995, ApJ 447, 23
\reference Oort J. H., 1983, ARA\&A 21, 373
\reference Peacock J.A., 1991, MNRAS 253, 1p
\reference Peacock J.A., Heavens A.F., 1990, MNRAS 243, 133
\reference Peacock J.A., Nicholson D., 1991, MNRAS 253, 307
\reference Peebles P.J.E., 1984, ApJ 284, 439
\reference Peebles P.J.E., 1993, Principles of Physical Cosmology, Princeton University Press
\reference Plionis M., 1993, SISSA preprint astro-ph/9312013
\reference Plionis M., Barrow J.D., Frenk C.S., 1991, MNRAS 249, 662
\reference Press W.H., Schechter P., 1974, ApJ 187, 425 
\reference Rhee G.F.R.N., Katgert P., 1987, A\&A 183, 217
\reference Rhee G.F.R.N., van Haarlem M.P., Katgert P., 1989, A\&AS 91, 513 
\reference Rood H.H., Page T.L., Kintner E.C., King I.R., 1972, ApJ 175, 627
\reference Salvador-Sol\'e E., Solanes J.M., 1993, ApJ 417, 427
\reference Saunders W., Frenk C., Rowan-Robinson M. et al., 1991, Nat 349, 32 
\reference Schaefer R.K., 1991, Int. J. Mod. Phys. A6, 2075
\reference Schaefer R.K., Shafi Q., 1993, Phys. Rev., D47, 1333
\reference Thomas P.S., Couchman H.M.P., 1992, MNRAS 257, 11
\reference Turner M.S., 1991, Phys. Scr. 36, 167
\reference Valdarnini R., Bonometto S.A., 1985, A\&A 146, 235
\reference van Kampen E., Rhee G.F.R.N., 1990, A\&A 237, 283 
\reference West M.J., 1994, MNRAS 268, 79
\reference White S.D.M.., Efstathiou G., Frenk C.S., 1993, MNRAS 262, 1023
\reference White S.D.M.., Frenk C.S., Davis M., Efstathiou G., 1987, ApJ 313, 505
\reference Zabludoff A.I., Geller M.J., Huchra J.P, Ramella M., 1993, AJ 106, 1301
\end{references}
\end{document}